\documentclass[iop,twocolappendix,numberedappendix,appendixfloats]{emulateapj}

\usepackage{apjfonts}

\usepackage{amsmath}
\usepackage[T1]{fontenc}

\newcommand{\Msun}{\,M_{\odot}}

\newcommand\ionn[2]{#1$\;${\scshape{#2}}}
\newcommand{\be}{\begin{equation}}
\newcommand{\ee}{\end{equation}}
\newcommand{\bea}{\begin{eqnarray}}
\newcommand{\eea}{\end{eqnarray}}
\def\kms{\ {\rm km\, s}^{-1}}

\def\logg{{\rm log}\,g}

\def\feh{\rm [Fe/H]}
\newcommand{\MS}{\texttt{MINESweeper}}

\shortauthors{CONROY ET AL.}
\shorttitle{Metallicities in the Stellar Halo}

\begin{document}


\title{Resolving the Metallicity Distribution of the Stellar Halo with
  the H3 Survey}

\author{Charlie Conroy\altaffilmark{1}, Rohan P. Naidu\altaffilmark{1},
  Dennis Zaritsky\altaffilmark{2}, Ana Bonaca\altaffilmark{1}, Phillip
  Cargile\altaffilmark{1}, Benjamin D. Johnson\altaffilmark{1}, Nelson
  Caldwell\altaffilmark{1}}

\altaffiltext{1}{Center for Astrophysics $\vert$ Harvard \& Smithsonian, Cambridge, MA, 02138, USA}
\altaffiltext{2}{Steward Observatory, University of Arizona, 933 North Cherry
  Avenue, Tucson, AZ 85721, USA}

\slugcomment{Submitted to ApJ}

\begin{abstract}

  The Galactic stellar halo is predicted to have formed at least
  partially from the tidal disruption of accreted dwarf galaxies.
  This assembly history should be detectable in the orbital and
  chemical properties of stars.  The H3 Survey is obtaining spectra
  for 200,000 stars, and, when combined with {\it Gaia} data, is
  providing detailed orbital and chemical properties of Galactic halo
  stars.  Unlike previous surveys of the halo, the H3 target selection
  is based solely on magnitude and {\it Gaia} parallax; the survey
  therefore provides a nearly unbiased view of the entire stellar halo
  at high latitudes.  In this paper we present the distribution of
  stellar metallicities as a function of Galactocentric distance and
  orbital properties for a sample of 4232 kinematically-selected halo
  giants to $100$ kpc.  The stellar halo is relatively metal-rich,
  $\langle$[Fe/H]$\rangle=-1.2$, and there is no discernable
  metallicity gradient over the range $6<R_{\rm gal}<100$ kpc.
  However, the halo metallicity distribution is highly structured
  including distinct metal-rich and metal-poor components at
  $R_{\rm gal}<10$ kpc and $R_{\rm gal}>30$ kpc, respectively.  The
  Sagittarius stream dominates the metallicity distribution at $20-40$
  kpc for stars on prograde orbits.  The {\it Gaia}-Enceladus merger
  remnant dominates the metallicity distribution for radial orbits to
  $\approx 30$ kpc.  Metal-poor stars with [Fe/H]$<-2$ are a small
  population of the halo at all distances and orbital categories.  We
  associate the ``in-situ'' stellar halo with stars displaying
  thick-disk chemistry on halo-like orbits; such stars are confined to
  $|z|<10$ kpc.  The majority of the stellar halo is resolved into
  discrete features in orbital-chemical space, suggesting that the
  bulk of the stellar halo formed from the accretion and tidal
  disruption of dwarf galaxies.  The relatively high metallicity of
  the halo derived in this work is a consequence of the unbiased
  selection function of halo stars, and, in combination with the
  recent upward revision of the total stellar halo mass, implies a
  Galactic halo metallicity that is typical for its mass.

\end{abstract}



\section{Introduction}
\label{s:intro}

The stellar halo provides a unique window into the assembly history of
our Galaxy.  The long dynamical times imply that the halo has not
undergone complete phase mixing, and therefore measurement of the
orbital and chemical properties of halo stars should enable a
reconstruction of the major events in the history of the Galaxy.

Early ideas concerning the formation of the stellar halo considered
both ``dissipative'' \citep{Eggen62}, and ``dissipationless''
\citep{Searle78} formation channels.  In modern terminology these are
referred to as ``in-situ'' and ``accretion'' (or ``ex-situ'')
channels.  In the former, halo stars are born within the Galaxy and
are by some dynamical mechanism heated to halo-like orbits
\citep[e.g.,][]{Abadi06, Zolotov09, Purcell10, Font11, Cooper15,
  Bonaca17}.  In the latter, hierarchical assembly in a cold dark
matter cosmology predicts that the halo was built at least in part by
the tidal disruption of smaller dwarf galaxies
\citep[e.g.,][]{Johnston96, Helmi99, Bullock05, Johnston08, Cooper10,
  Font11}.

In principle the combined orbital and chemical properties of stars
should provide a powerful approach to understanding the origin of the
halo.  For example, such data should enable the categorization of
in-situ and accreted stars as a function of distance, metallicity,
etc.  A major goal is to identify the number of significant events
that contributed to the accreted halo, estimate their progenitor
masses and orbital properties, and ultimately reconstruct the build-up
of the stellar halo.  This field has a long and rich history of using
the chemical-orbital properties of stars to study the origin of the
halo \citep[e.g.,][]{Sommer-Larsen90, Ryan91, Majewski92, Zinn93,
  Carney94, Chiba00, Carollo07, Bell08, Morrison09, Carollo10,
  Bonaca17, Helmi18, Belokurov18, Lancaster19, Iorio19}.

To-date, nearly all observational work on the stellar halo has
employed tracers that are biased with regards to metallicity.  These
biases ultimately stem from the fact that halo stars are rare and
generally more metal-poor relative to the disk, combined with a desire
to make efficient use of spectroscopic resources.  Prior to {\it
  Gaia}, the most efficient way to separate halo from disk stars was
to select stars with low metallicities.  This bias can arise at two
distinct stages in the analysis: first in the selection of targets for
spectroscopic followup, and second via the identification of halo
stars from the final sample.  For example, the SDSS calibration stars
used by \citet{Carollo07, Carollo10} to study the stellar halo were
selected on the basis of their blue colors.  The SDSS SEGUE sample of
K giants, which has been used to study the halo to great distances
\citep[e.g.,][]{Xue15, Das16}, was selected for spectroscopic
follow-up on the basis of a complex set of color-cuts that favors low
metallicities.  Photometric metallicities of F/G turnoff stars are
another popular method for studying the stellar halo.  However, such
samples are also constructed on the basis of color-cuts, and favor
lower metallicity stars \citep[e.g.,][]{Ivezic08, Sesar11, Zuo17}.
Rare populations such as RR Lyrae and blue horizontal branch stars are
another popular tracer of the halo, in part because they are standard
candles \citep[e.g.,][]{Cohen17, Lancaster19, Iorio19}.  However,
these populations also preferentially trace metal-poor stars.  Biases
incurred by using these populations to study the halo are very
difficult to overcome without near-perfect knowledge of the underlying
population and what fractions of stars were and were not included in
the sample.  These different observational methods have resulted in
sometimes conflicting conclusions regarding the chemical-orbital
structure of the stellar halo.

Thankfully, the observational landscape is rapidly improving on
multiple fronts.  {\it Gaia} has measured proper motions and
parallaxes for $>1$ billion stars to $G\approx20$ \citep{GaiaDR2}.
The majority of halo stars are too distant to have precise parallaxes,
and are too faint to have a measured {\it Gaia} radial velocity.  To
complement {\it Gaia} in the halo, we are undertaking the H3 Stellar
Spectroscopic Survey of high latitude fields \citep{Conroy19a}.  The
survey is delivering radial velocities, metallicities, and
spectrophotometric distances for 200,000 parallax-selected stars.  The
key novelty of the H3 survey is a very simple selection function.  H3
combined with {\it Gaia} is providing, for the first time, an unbiased
view of the stellar halo to distances of $100$ kpc.

In this paper we present the metallicity distribution of the stellar
halo as a function of Galactocentric radius, vertical position
(relative to the disk), and orbital properties.  These results are
used to understand the origin(s) of the Galactic stellar halo.


\section{Data}
\label{s:data}

The H3 Survey \citep{Conroy19a} is collecting spectra for 200,000
stars in high latitude fields.  The survey footprint covers
Dec.$>-20^{\circ}$ and $|b|>30^{\circ}$.  The selection function is
very simple and consists of a magnitude limit of $r<18$ and a parallax
selection of $\pi<0.5$ mas.  Data collection began before {\it Gaia}
DR2 was available, and so before the parallax selection could be made
we obtained spectra for all stars with $g-r<1.0$.  19,300 stars were
observed in this way (21\% of the current sample).  This color-cut is
very mild -- in the parallax-selected sample only 5\% of stars have
$g-r>1$.  None of the results presented below change if the early data
are removed from the analysis.

The survey employs the medium-resolution Hectochelle spectrograph
\citep{Szentgyorgyi11} on the MMT.  Hectochelle uses a robotic fiber
positioning system \citep{Fabricant05}, enabling the placement of 240
fibers over a $1^{\circ}$ diameter field of view.  The instrument is
configured to deliver $R\approx23,000$ spectra over the wavelength
range $5150$\AA$-5300$\AA.  As of June 2019 the survey has collected
89,000 spectra over 469 fields restricted to $|b|>40^{\circ}$.
Details of the survey design and data quality can be found in
\citet{Conroy19a}.

Stellar parameters, including radial velocities, spectrophotometric
distances, and abundances ([Fe/H] and [$\alpha$/Fe]) are measured
using the \MS\, program \citep{Cargile19}.  The dominant $\alpha$
element in the H3 wavelength region is the \ionn{Mg}{i} triplet, so
[$\alpha$/Fe] is mostly tracing [Mg/Fe].  Briefly, \MS\, combines
spectral libraries and stellar isochrones to simultaneously fit for
stellar parameters along with distance and redenning.  \MS\, uses a
Bayesian framework to fit the continuum-normalized spectrum and the
broadband photometry (including Pan-STARRS, {\it Gaia}, 2MASS, WISE,
and SDSS where available).  {\it Gaia} parallaxes are used as a prior.
Uncertainties on radial velocities determined from repeat observations
are very small ($<1\kms$).  Derived distances have a typical
uncertainty of $\approx10$\% for giants.  {\it Gaia} DR2 proper
motions for the H3 stars have median SNR of 25 and 33 for the R.A. and
Dec. components.  For these two components, 86\% and 91\% of the
sample have SNR$>5$.

For calibration purposes the H3 Survey has obtained spectra of stars
in the globular clusters M92, M3, M13, M71, and M107, and the open
cluster M67.  Together, these clusters span a range in metallicities
from [Fe/H]$=-2.3$ to $+0.0$.  \citet{Cargile19} demonstrated that
\MS\, accurately determines distances, stellar parameters,
metallicities and abundances of the cluster stars.  The literature
values for most globular clusters show a scatter of $\approx0.1$ dex
and \MS\, often returns metallicities at the upper end of this range.
These tests lead us to conclude that \MS-derived metallicities are
accurate to $\lesssim 0.1$ dex.  This issue is further discussed in
Section \ref{s:lim} and Appendix \ref{s:comp}.

In this paper we use a high-quality subset of the full dataset.  Stars
are selected that have a quality flag$=0$ (removing bad data and very
poor fits; $\approx1$\% of the sample).  We also place a limit on the
median signal-to-noise ratio across the H3 spectrum such that SNR$>3$,
which results in a higher purity sample of metallicities.  We remove
the small number of blue horizontal branch stars that were explicitly
targeted, since they have unreliable metallicities, and a small number
of stars with very large tangential velocities ($v_T>700\kms$); visual
inspection indicates that their stellar parameters are wrong and are
therefore at incorrect distances.  Giants with a derived rotational
velocity $>5\kms$ are removed, as visual inspection indicates that
these are dwarf stars being erroneously fit as a broadened giant.
This affects 1\% of the current sample and will be dealt with in a
future version of the catalog by introducing a $\logg-$dependent prior
on the rotational broadening.  These cuts leave 63,694 stars.

The spectrophotometric distances, radial velocities, and {\it Gaia}
proper motions are then used to derive a variety of quantities
including projections of the angular momentum vector onto the
Galactocentric coordinate system \citep[assuming the local standard of
rest from][]{Schonrich10}.  Here we use the $z-$component of the
angular momentum vector, $L_z$, as a way to group stars by their
orbital properties (in our right-handed coordinate system, prograde
stars have $L_z<0$ and retrograde stars have $L_z>0$).  

We focus in this paper on kinematically-selected halo giants.
Specifically, we require $|V-200|>180 \kms$ where $V$ is the 3D
velocity, and $\logg<3.5$.  The kinematic selection efficiently
removes stars on disk-like orbits \citep[e.g.,][]{Venn04, Nissen10},
and results in a sample of 14,152 stars.  The giant selection ensures
that the sample is not dominated by nearby halo dwarf stars, and
results in a final sample of 4232 stars.

\begin{figure}[!t]
\center
\includegraphics[width=0.47\textwidth]{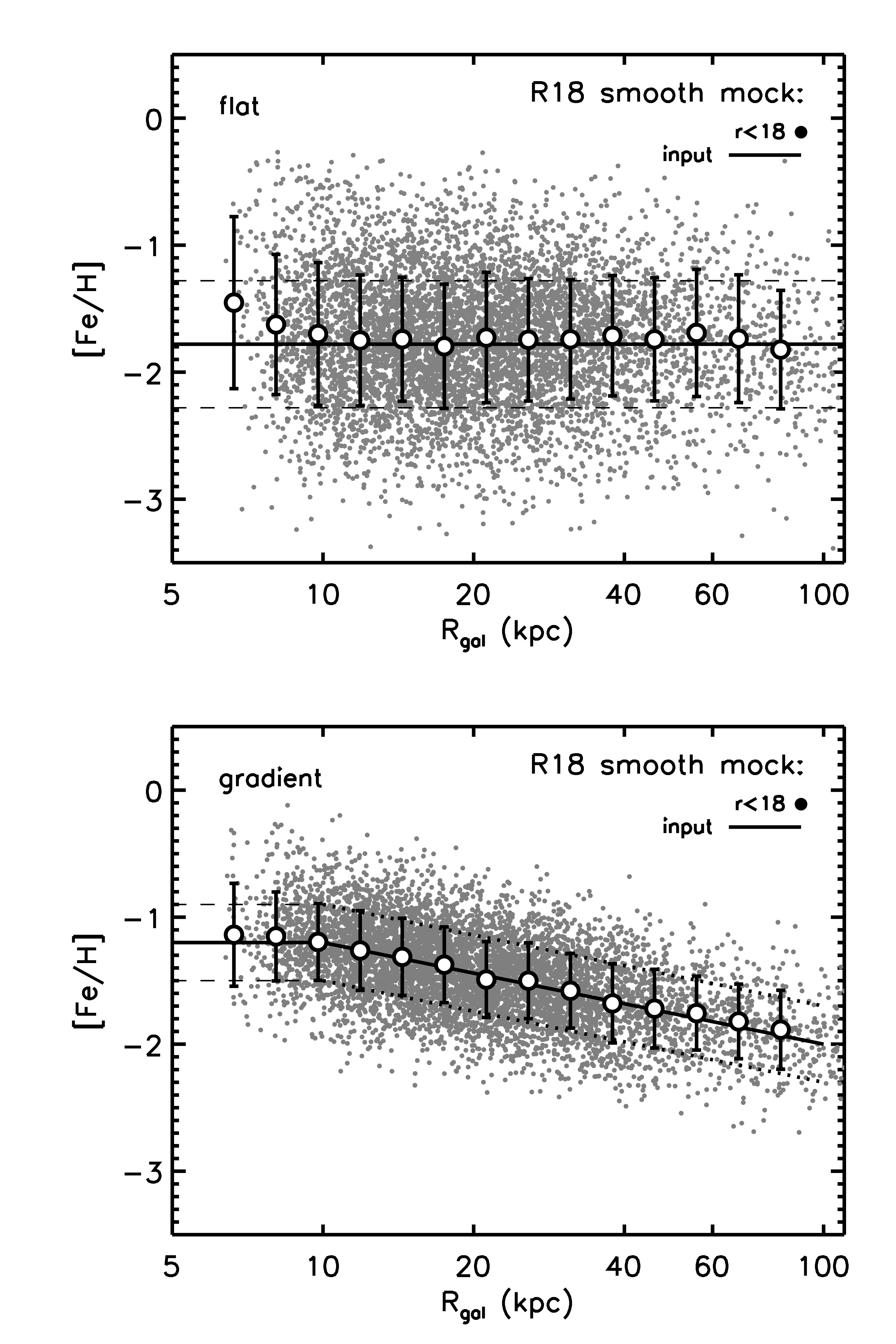}
\vspace{0.1cm}
\caption{Effect of the H3 selection function on the recovered
  metallicity profile in the halo.  Top and bottom panels show two
  models for the input metallicity profile of the R18 mock stellar
  halo.  Small points show the metallicities of kinematically-selected
  halo stars drawn from the R18 mock catalog subject to the H3
  selection function ($r<18$, $\pi<0.5$) and the H3 window function.
  Large open symbols and errors show the median and standard deviation
  in radial bins.  Solid and dashed lines show the true input
  metallicity distribution of the R18 halo (mean and scatter) for the
  two model metallicity profiles.  Agreement between the open symbols
  and lines indicates that the H3 selection function does not impose a
  metallicity bias as a function of radius.}
\label{fig:mock}
\end{figure}

\begin{figure}[!t]
\center
\includegraphics[width=0.47\textwidth]{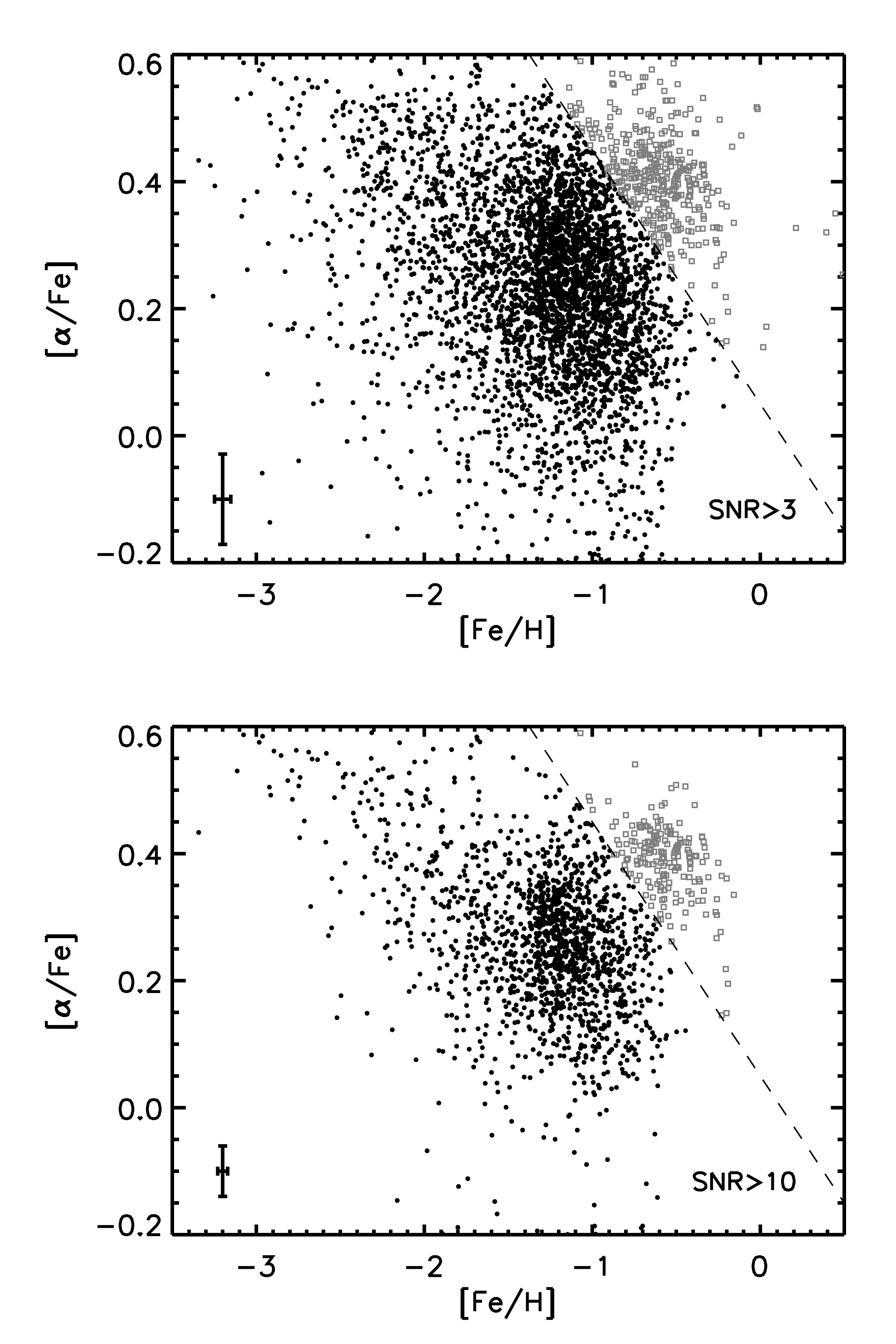}
\vspace{0.1cm}
\caption{[$\alpha$/Fe] vs. [Fe/H] for kinematically-selected halo
  giants in the H3 Survey.  Top panel shows the fiducial sample with
  spectral SNR $>3$.  In the bottom panel we show a subset of the data
  with SNR $>10$ where the various subpopulations are even more
  clearly visible.  Stars above the dashed line have thick disk
  chemistry and so are associated with the in-situ stellar halo.
  Median uncertainties are shown in the lower left corner of each
  panel.}
\label{fig:h3afe}
\end{figure}

\begin{figure*}[!t]
\center
\includegraphics[width=0.9\textwidth]{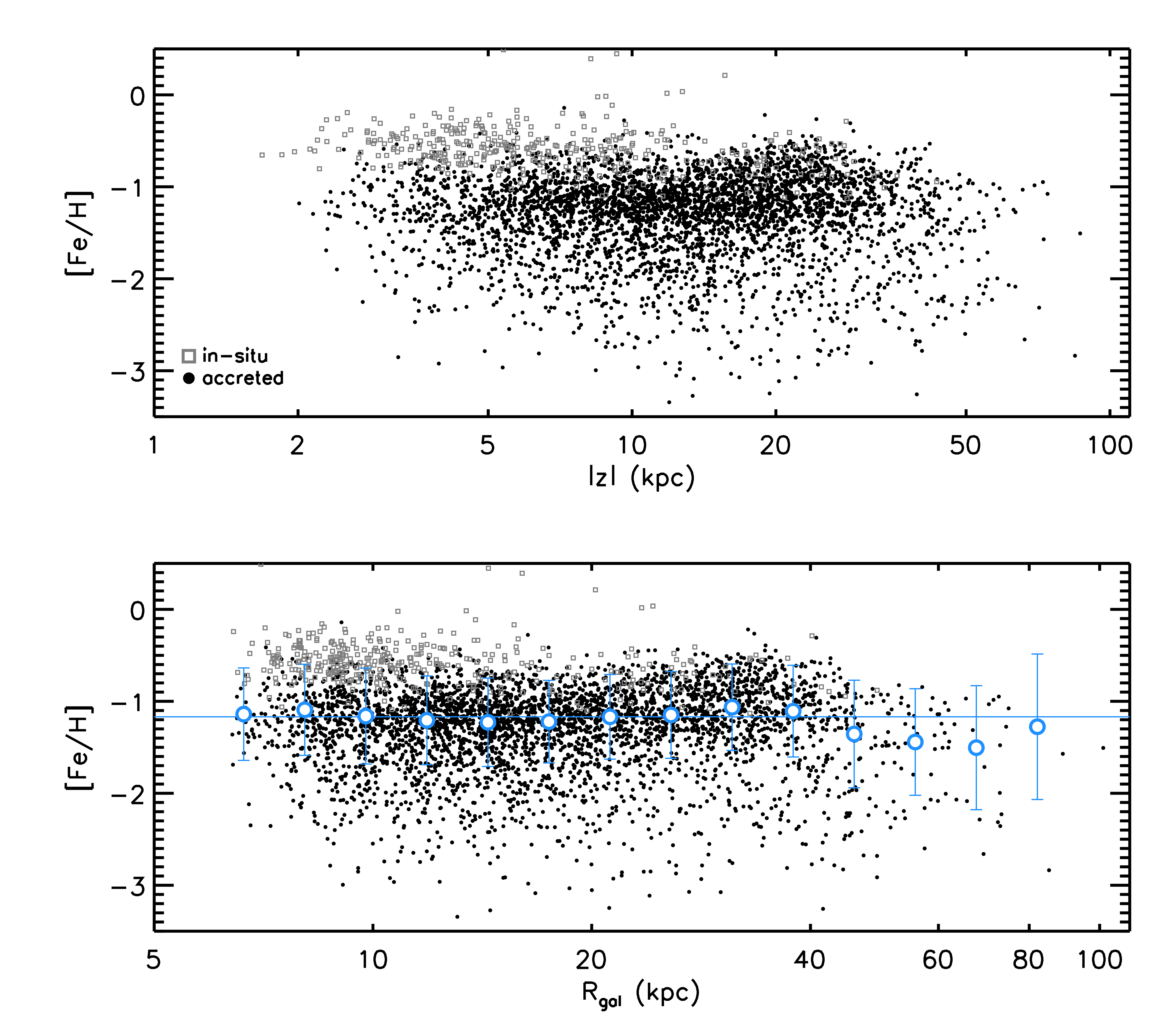}
\vspace{0.1cm}
\caption{Stellar metallicity vs. distance from the Galactic plane (top
  panel) and Galactocentric radius (bottom panel) for
  kinematically-selected halo stars from the H3 Survey.  Grey points
  have thick disk chemistry as defined in Figure \ref{fig:h3afe} and
  are therefore defined as the in-situ stellar halo. In the bottom
  panel, the mean and scatter is shown in blue as a function of
  radius.  The overall profile is remarkably flat with
  $\langle \feh \rangle=-1.2$, although there are clearly multiple
  distinct populations. The median measurement uncertainty on [Fe/H]
  is 0.05 dex.}
\label{fig:zRgal}
\end{figure*}

\begin{figure*}[]
\center
\includegraphics[width=0.95\textwidth]{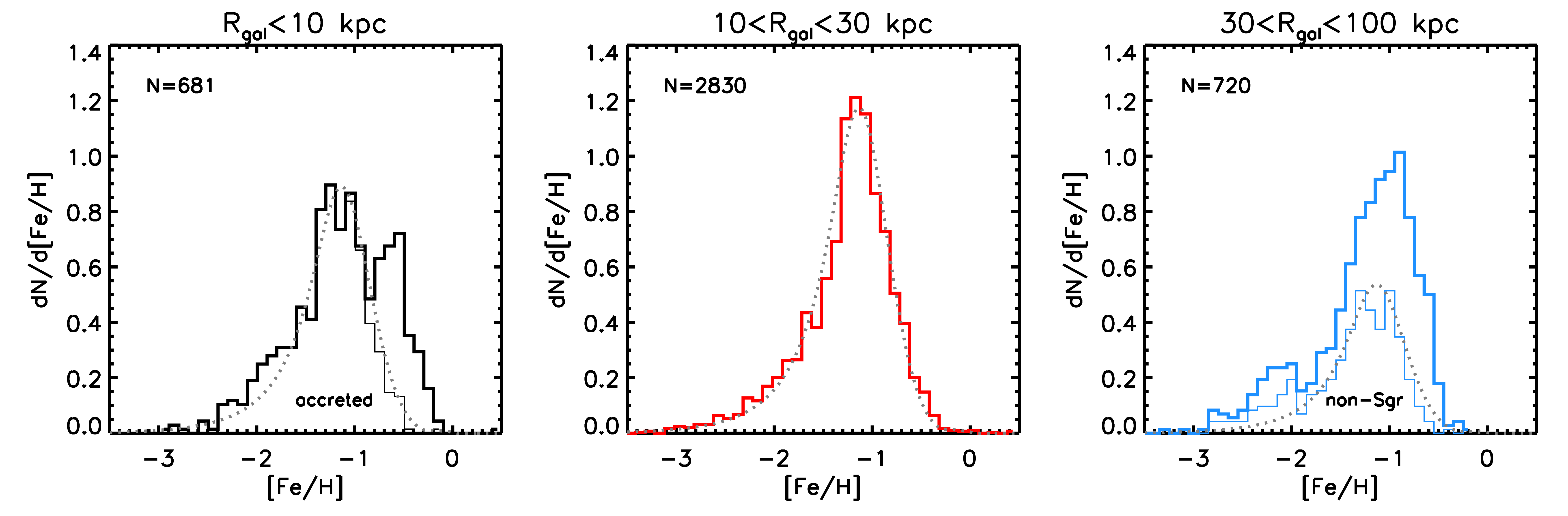}
\vspace{0.1cm}
\caption{Metallicity distribution functions (MDFs) in three radial bins for
  kinematically-selected halo giants.  The grey dotted line is an
  analytic chemical evolution model fit to the distribution in the
  middle panel.  In the left panel, we also show the MDF excluding the
  in-situ halo stars (thin solid line labeled ``accreted''), while in
  the right panel we show the MDF excluding Sagittarius stream stars.
  In the left and right panels the dotted line is scaled to the number
  of stars in the accreted and non-Sgr components, respectively.  The
  bin width used to compute the histograms is 0.1 dex.}
\label{fig:mdf}
\end{figure*}


\section{Test of the Selection Function with Mock Data}
\label{s:mock}

In this section we use a mock star catalog of the Galaxy in order to
investigate the impact of our selection function on the inferred
global properties of the halo.

\citet[][R18]{Rybizki18} present a mock Galaxy tailored to {\it
  Gaia}-like data.  The mock catalog is based on the Galaxia synthetic
Galaxy \citep{Sharma11} and incorporates the major stellar components
of the Galaxy including thin and thick disks, a bulge and a stellar
halo.  R18 adopt the {\it Gaia} DR2 error model for uncertainties on
proper motions and include a realistic 3D dust extinction map.

The default stellar halo is uniformly old (13 Gyr) and has a
metallicity distribution function that is Gaussian with a mean of
[Fe/H]$=-1.78$, a dispersion of $\sigma_{\rm [Fe/H]}=0.5$, and no
gradient with Galactocentric radius.  We have also explored a modified
mock catalog in which the stellar halo has a metallicity gradient that
is linear in log($R_{\rm gal}$) from $\feh=-1.2$ to $\feh=-2.0$ over
the range $10-100$ kpc.  The profile is flat at $<10$ kpc and $>100$
kpc.  We have recomputed photometry self-consistently for both
versions of the mock catalog using the \texttt{MIST} isochrones and
bolometric correlations \citep{Choi16}.

We have taken the R18 mock and made several modifications in order to
approximate the H3 Survey data.  We impose the H3 selection function
($r<18$ and $\pi<0.5$ mas) and the H3 window function (keeping only
stars that lie within the FOV of our observations).  After applying
the selection function, we randomly select a maximum of 200 stars
within each FOV, as only $\approx200$ stars are assigned fibers per
pointing.  We then apply the same kinematic halo selection as
discussed in Section \ref{s:data} and select giants with $\logg<3.5$.
Finally, we impose a 10\% fractional uncertainty on the distances.
This is the median distance uncertainty for the giants in H3
\citep{Conroy19a}.

With these H3-like mock catalogs we are now in a position to assess
the H3 selection function on the metallicity profile in the halo.  The
H3 selection function is very simple, but one could imagine that even
a magnitude selection might result in a bias owing to the fact that
the most luminous giants are brighter in the $r-$band at lower
metallicities.

We test this effect in Figure \ref{fig:mock}, which shows the
distribution of metallicities vs. radius for the
kinematically-selected halo giants in the two versions of the R18 mock
catalog.  The top panel shows the default R18 stellar halo metallicity
model: a flat profile in radius.  The bottom panels shows the model
that has a gradient from $10-100$ kpc.  Median metallicities and
$1\sigma$ scatter values are computed in radial bins (points with
error bars), and compared to the true underlying distribution (solid
and dotted lines).  The excellent agreement implies that the H3
selection function does not impart a bias in the recovered metallicity
gradient.

\begin{figure*}[!t]
\center
\includegraphics[width=0.95\textwidth]{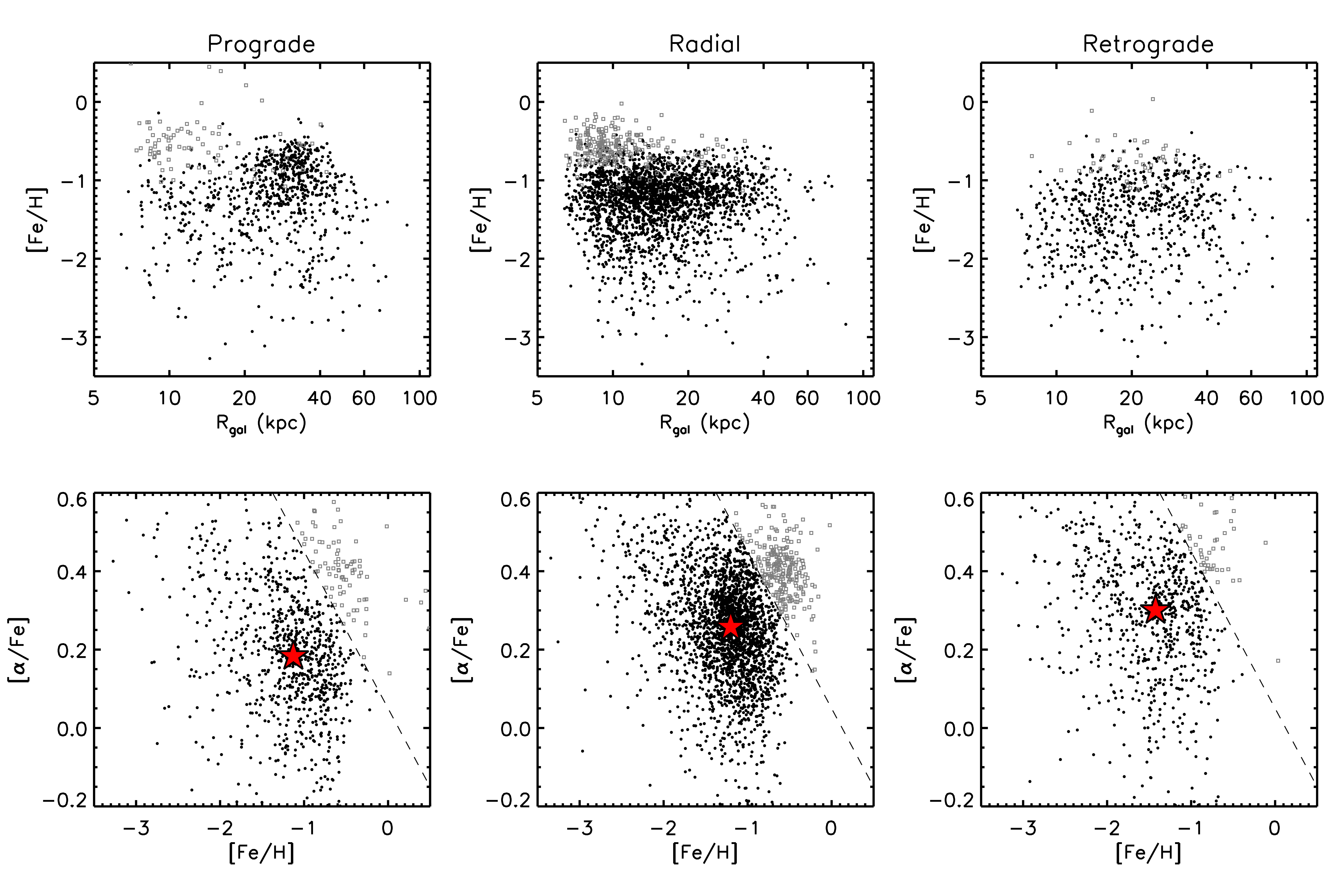}
\vspace{0.1cm}
\caption{Top Panels: Metallicity vs. Galactocentric radius separated
  according to the $z-$component of the angular momentum (prograde,
  radial, and retrograde in the left, middle and right panels).
  Bottom panels: Distribution of stars in [$\alpha$/Fe] vs. [Fe/H].
  Only kinematically-selected halo stars are shown.  Grey points lie
  above the dashed lines in the bottom panels and mark the in-situ
  halo stars.  The Sagittarius stream is prominent in the left panels.
  The {\it Gaia}-Enceladus remnant dominates the middle panels along
  with a metal-rich, $\alpha-$rich population of in-situ stars.  The
  right panels is perhaps dominated by the Sequoia remnant.  In the
  lower panels, the red star marks the median values of $\feh$ and
  [$\alpha$/Fe] for the accreted stars (black points).}
\label{fig:h3lz}
\end{figure*}

\begin{figure*}[!t]
\center
\includegraphics[width=0.95\textwidth]{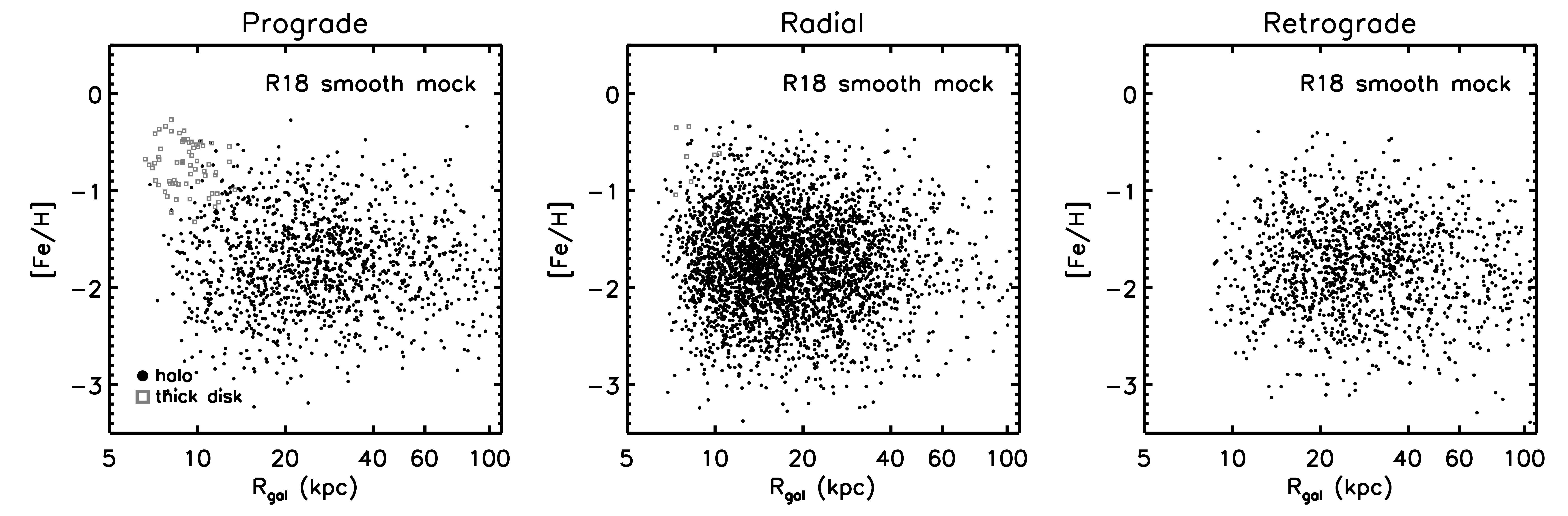}
\vspace{0.1cm}
\caption{As in Figure \ref{fig:h3lz}, now showing
  kinematically-selected halo stars from the R18 smooth mock catalog.
  Black points indicate true halo stars while grey points belong to
  the thick disk.  In this model the halo population is intrinsically
  smooth with a power-law density profile and a flat metallicity
  profile.  The presence of grey points in the left panel means that
  thick disk stars are entering into the kinematic halo selection.
  These stars are not present in significant numbers amongst the
  radial orbits (middle panel), in stark contrast to the data (Figure
  \ref{fig:h3lz}, middle panel).}
\label{fig:r18lz}
\end{figure*}

\begin{figure*}[!t]
\center
\includegraphics[width=0.95\textwidth]{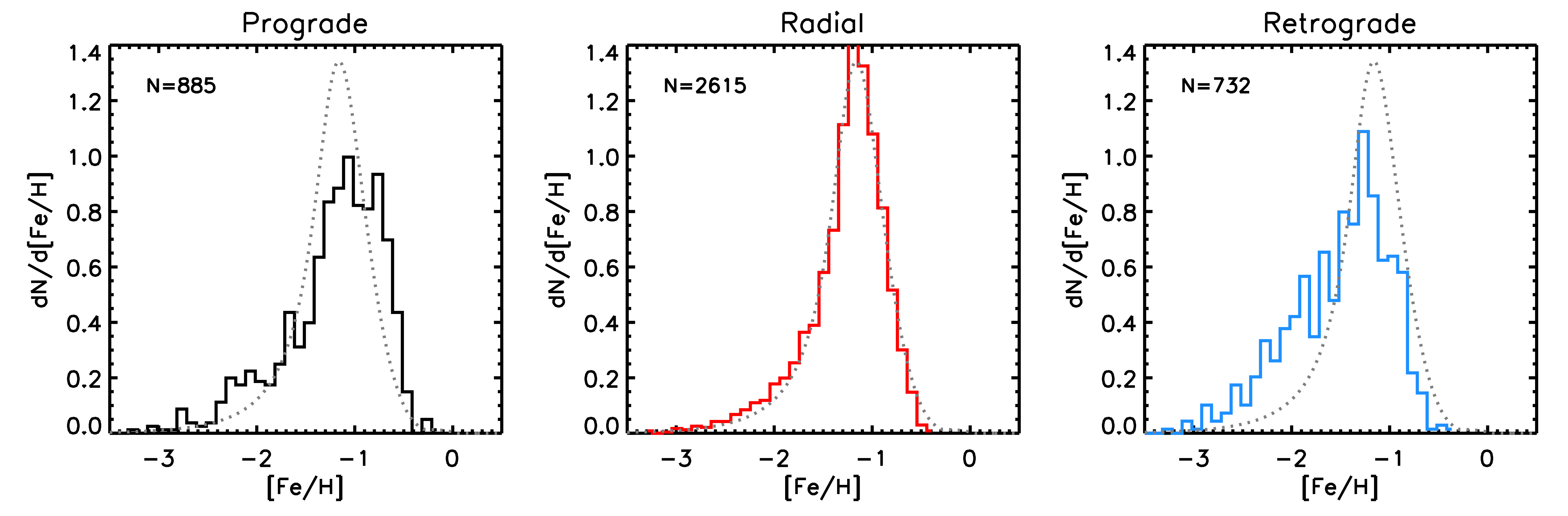}
\vspace{0.1cm}
\caption{Metallicity distribution functions (MDFs) of H3 stars shown
  for prograde, radial, and retrograde orbits.  Here we have removed
  the in-situ halo stars (those above the dashed lines in the lower
  panels of Figure \ref{fig:h3lz}).  The dotted line is a simple
  chemical evolution model fit to the MDF in the middle panel and
  replicated in other panels for comparison purposes.  There are
  multiple chemically-distinct stellar populations amongst the
  prograde and retrograde stars.  The population of stars on radial
  orbits are consistent with arising from a single stellar
  population.}
\label{fig:mdflz}
\end{figure*}


\vspace{2cm}

\section{Results}
\label{s:res}

\subsection{The Halo Metallicity Profile}
\label{s:fehprof}

We begin with an overview of the abundance patterns of the
kinematically-selected halo giants.  Figure \ref{fig:h3afe} shows
[$\alpha$/Fe] vs. [Fe/H] for the main sample (top panel), and for a
high SNR subset (bottom panel).  There are several distinct
populations in this diagram, though we draw attention to the stars
above the dashed line.  These stars have thick disk chemistry and yet
are on halo-like orbits \citep[see also][]{Bonaca17}.  We identify
such stars with the ``in-situ'' stellar halo and discuss their
location in various diagrams below.

Figure \ref{fig:zRgal} shows the metallicity profile of
kinematically-selected halo giants from the H3 Survey as a function of
distance from the Galactic plane (top panel) and Galactocentric radius
(bottom panel).  In the bottom panel we also show median metallicities
and $1\sigma$ scatter in radial bins.  The median metallicity of the
entire sample is $\langle \feh \rangle=-1.2$ and is shown as a solid
line.  Stars with thick disk chemistry are shown as grey points.

There are several important features in Figure \ref{fig:zRgal}.  The
overall metallicity profile is remarkably flat across the entire range
from $\approx6-100$ kpc.  There is marginal evidence for a lower mean
metallicity beyond $\approx50$ kpc, but there are too few stars in the
current data to draw strong conclusions.  Importantly, the average
metallicity is considerably more metal-rich than most previous work.
We return to this point in Section \ref{s:disc}.  There are two
populations that are more metal-rich than the rest of the halo.  The
first is at $|z|<5$ kpc and is associated with the in-situ halo (i.e.,
stars having thick disk-like chemistry).  The second metal-rich
component is at $20\lesssim R_{\rm gal}\lesssim 40$ kpc and is
associated with the Sagittarius stream.  Finally, there is a clear
metal-poor component ([Fe/H]$\lesssim-2$) that extends to $\approx100$
kpc.

The metallicity distribution function (MDF) is shown in Figure
\ref{fig:mdf} in three radial bins.  At $R_{\rm gal}<10$ kpc one
clearly sees evidence for two distinct populations, including a main
population with a mean metallicity of $\feh=-1.2$ and a secondary
metal-rich population.  Removing stars belonging to the in-situ halo
as defined in Figure \ref{fig:h3afe} results in an MDF with a single
peak at $\feh=-1.2$, shown as the thin line in Figure \ref{fig:mdf}.

At $10<R_{\rm gal}<30$ kpc the distribution is consistent with a
single population with a mean metallicity of $\feh=-1.2$.  To explore
this further, we fit the MDF with a simple chemical evolution model.
\citet{Kirby11} modelled the MDFs of a sample of dwarf galaxies using
a variety of simple chemical evolution models.  They found that the
``Best Accretion Model'' of \citet{Lynden-Bell75} overall performed
well in reproducing the observed MDFs.  In this model the gas mass has
a non-linear dependence on the stellar mass, quantified by the
parameter $M$ which is the ratio between the final and initial mass of
the system.  We use this model and fit its two free parameters (the
yield, $p=0.08$, and $M=2.1$) to the data in the middle panel of
Figure \ref{fig:mdf}.  The result is shown as a dotted line, and is a
good fit to the observed MDF, suggesting that one population dominates
this radial range.

In the third panel of Figure \ref{fig:mdf} we show the MDF for
$30<R_{\rm gal}<100$ kpc.  There are at least two distinct populations
based on the MDF alone: a metal-rich population at $\feh\approx-1$
and a metal-poor population at $\feh\approx-2.1$.  We have identified
stars likely belonging to the Sagittarius stream according to their
distribution in $L_y-L_z$ space.  Specifically, stars with
$L_y<-L_z-3\times10^3 \, {\rm kpc} \kms$ are selected as Sagittarius
stars (see Johnson et al. in prep for details).  Removing these stars
from the MDF results in the thin line in Figure \ref{fig:mdf}.  Even
after removing Sagittarius there are clearly at least two distinct
chemical populations. 

\subsection{Halo Metallicities vs. Orbital Properties}

In this section we investigate the dependence of halo metallicities on
orbital properties.  In particular, we focus on the $z-$component of
the orbital angular momentum, $L_z$, and we define three groupings of
stars: prograde ($L_z<-5\times10^2\, {\rm kpc} \kms$), retrograde
($L_z>10\times10^2\, {\rm kpc} \kms$), and radial orbits
($-5\times10^2 <L_z<10\times10^2 \,{\rm kpc} \kms$).  The quantitative
selection was chosen based on the distribution of stars in $E-L_z$
space, where $E$ is the total energy of the orbit (the distribution of
H3 stars in $E-L_z$ will be presented in Naidu et al. in prep).  As
shown in previous work \citep[e.g.,][]{Helmi18, Belokurov18} and with
H3 data in Naidu et al. (in prep.), there is a population of stars on
strongly radial orbits that cluster in the radial orbit selection
region we have outlined.  This population has been referred to in the
literature as {\it Gaia}-Enceladus \citep{Helmi18} and the {\it
  Gaia}-Sausage \citep{Belokurov18}, with slight differences in how
the population is defined in each case.  There is a slight asymmetry
in the distribution, which led us to impose an asymmetric selection in
$L_z$.

\begin{figure*}[!t]
\center
\includegraphics[width=0.95\textwidth]{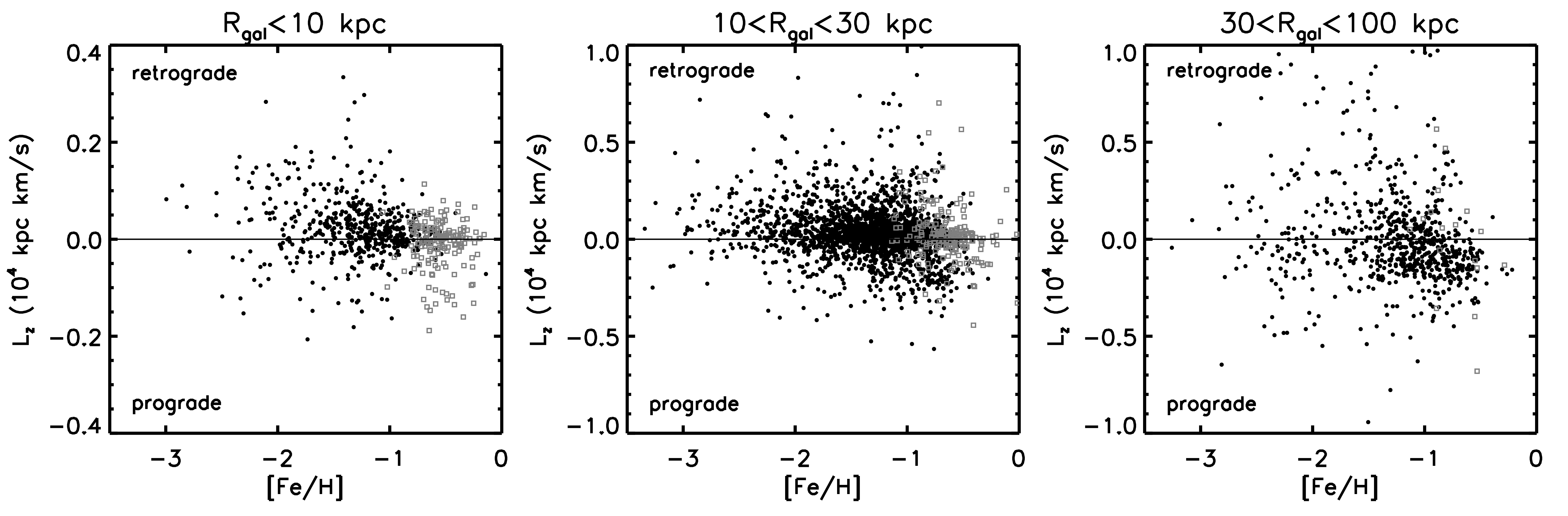}
\vspace{0.1cm}
\caption{$L_z$ vs. [Fe/H] for kinematically-selected halo giants from
  the H3 Survey.  Stars are grouped in three radial bins:
  $R_{\rm gal}<10$ kpc, $10<R_{\rm gal}<30$ kpc, and
  $30<R_{\rm gal}<100$ kpc.  Grey points mark the in-situ halo stars
  defined in the bottom panels of Figure \ref{fig:h3lz}.}
\label{fig:h3fehlzR}
\end{figure*}

In Figure \ref{fig:h3lz} we show the metallicities of
kinematically-selected halo giants as a function of orbital
properties.  In the top panels we show $\feh$ vs. Galactocentric
radius and in the bottom panels we show $\feh$ vs. [$\alpha$/Fe].  

The top panels display a wealth of structure.  The prograde-halo
population is dominated by a relatively metal-rich feature at
$20<R_{\rm gal}<40$ kpc.  This is the Sagittarius stream, and will be
discussed in detail in Johnson et al. (in prep.).  There is also a
distinct metal-poor population at $\feh\approx-2$.  The radial-halo
population is composed of two principle populations.  The dominant
population is at $\feh\approx-1.2$ and extends to $\approx30$ kpc.
This is the {\it Gaia}-Enceladus merger remnant.  There is also a
metal-rich ($\feh>-1$) population confined to $R_{\rm gal}\lesssim20$
kpc ($|z|\lesssim10$ kpc; grey points) that we associate with the
in-situ halo.  The retrograde-halo population is on average more
metal-poor than the other two orbital groupings.  There is a
population at $\feh\approx-1.2$ that clearly has a more extended
distribution in Galactocentric radius compared to the halo stars on
radial orbits.  In addition, there is a relatively prominent
metal-poor population that is also quite extended in radius.

In the bottom panels of Figure \ref{fig:h3lz} one sees systematic
variation in [$\alpha$/Fe] with orbital properties.  The prograde halo
(dominated by the Sagittarius stream) is relatively $\alpha-$poor; the
radial group (dominated by {\it Gaia}-Enceladus) is more
$\alpha-$rich, while the retrograde group is the most $\alpha$-rich
and metal-poor.

In Figure \ref{fig:r18lz} we show the metallicity distributions of
kinematically-selected halo stars separated by orbital properties for
the fiducial R18 smooth mock catalog.  As a reminder, this mock
catalog was generated assuming intrinsically smooth distributions of
the thin and thick disks and a single-component smooth stellar halo.
The mock dataset has the H3 selection function applied and a realistic
error model for all relevant parameters.  True halo stars are shown as
black symbols, while thick disk stars are shown in grey.  As expected,
the distribution of stars in Figure \ref{fig:r18lz} is smooth, and
there is no discernable orbital dependence of the halo population.
Within the prograde group there is small population of thick disk
stars at $R_{\rm gal}<10$ kpc.  This population is not a consequence
of the error model in the mock data but instead seems to simply be the
tail of the thick disk distribution.  Note that thick disk stars are
not present in significant numbers amongst the radial orbit group, in
contrast to the data.

Figure \ref{fig:mdflz} shows the MDF of kinematically-selected halo
giants in the three orbital groups.  In this figure we have removed
the metal-rich and $\alpha-$rich stars (the in-situ stars) that lie
above the dashed lines in the lower panels of Figure
\ref{fig:h3lz}. The dotted line is the Best Accretion chemical model
fit to the data in the middle panel, and reproduced in the other
panels for comparison purposes.  The MDFs in Figure \ref{fig:mdflz}
add support to the interpretation of Figure \ref{fig:h3lz} discussed
above.  In particular, the prograde and retrograde groups clearly show
at least two distinct populations (even after removing the in-situ
component), while the radial group appears to be dominated by a single
population.

In Figure \ref{fig:h3fehlzR} we show the distribution of halo stars in
$L_z$ vs. [Fe/H].  The H3 sample is grouped into three radial bins:
$R_{\rm gal}<10$ kpc, $10<R_{\rm gal}<30$ kpc, and
$30<R_{\rm gal}<100$ kpc.  We also mark in grey the in-situ halo
stars.  Note that stars at greater distances naturally occupy a wider
range in $L_z$ values, which explains why stars at smaller
$R_{\rm gal}$ are confined to a relatively narrow range in $L_z$.

There are multiple distinct populations evident in Figure
\ref{fig:h3fehlzR}.  The Sagittarius stream comprises the prograde
metal-rich region, while the {\it Gaia}-Enceladus remnant dominates
the radial $L_z\sim0$ region for $R_{\rm gal}<30$ kpc.  At $\feh<-2$
there is a strongly retrograde population at $R_{\rm gal}<30$ kpc -
the nature of the metal-poor population at greater distances is
unclear.  Finally, there is a hint of a retrograde population at
$-2<\feh<-1$, which may be associated with the Sequoia merger event
\citep{Matsuno19, Myeong19a}.


\section{Discussion}
\label{s:disc}

\subsection{Caveats and Limitations}
\label{s:lim}

In \citet{Cargile19} we provided multiple tests of the H3 stellar
parameters, including comparison to five globular clusters and M67.
The overall metallicities and [$\alpha$/Fe] values were in good
agreement with literature estimates.  However, in some cases the
derived metallicities were $0.05-0.1$ dex higher than previous work.
It is possible that our overall metallicity scale is therefore
slightly too high, though the magnitude of the effect is likely less
than 0.1 dex.  We explore this issue further in Appendix \ref{s:comp},
where H3 metallicities are compared to APOGEE, LAMOST, and SEGUE
metallicities for stars in common between the surveys.

The current H3 footprint is restricted to $|b|>40^\circ$ with many
more fields in the Northern hemisphere \citep{Conroy19a}.  Upon
completion of the survey the footprint will sparsely cover
$\approx35$\% of the sky.  A full accounting of the stellar halo must
take the survey footprint into account.  For example, structure that
is confined near the disk plane will be missed in a high latitude
survey.  Special care must also be given to halo structure that is at
least somewhat coherent on the sky (such as Sagittarius).  We have not
attempted any such corrections in the present work, so quantitative
determination of the mass fraction in various halo structures must be
viewed as preliminary.

Finally, we caution that our spectroscopic survey is able to detect as
coherent structures in phase space only those systems that had a
relatively high progenitor stellar mass.  We can estimate a rough mass
limit as follows.  The H3 Survey has covered 370 sq. deg. to-date,
which represents $\approx1$\% of the sky.  We have focused here on
giants with $\logg<3.5$.  Such stars comprise $\approx0.3$\% of the
stars in an old stellar population \citep[estimated from the
\texttt{MIST} isochrones assuming a Kroupa IMF;][]{Choi16}.  If we
optimistically assume that we have obtained a spectrum for every giant
within each field of view, then our sampling rate is approximately
1:300,000.  This number can be estimated another way: if we assume
that there are $\approx10^9$ stars in the stellar halo
\citep{Deason19}, and our current sample of halo giants contains 4232
stars, then the sampling rate is 1:250,000 -- reasonably close to the
previous estimate.  If one assumes that 100 stars are required in
order to identify a cold feature in phase space, then our sensitivity
limit is in the range of $M_{\ast}\approx3\times10^7\Msun$.  In
detail, this limit will be lower for systems that disrupted nearer to
the solar neighborhood compared to more distant systems.  This is due
to the fact that our magnitude limit corresponds to relatively more
luminous stars at greater distances, and such stars are an
intrinsically smaller fraction of the underlying population.  The main
point is that any survey of the halo will only be sensitive to
disrupted systems above a certain mass threshold, and this threshold
must be taken into account when interpreting the results.

\subsection{Comparison to Previous Work}

There is a large body of work exploring the metallicity and orbital
properties of the stellar halo.  When comparing to previous work,
several issues must be kept in mind: 1) The selection of stars
entering the spectroscopic catalog frequently is strongly biased
toward a particular stellar population.  For example, spectra for
Sagittarius stream stars have often been obtained for stars satisfying
the M giant color-cuts of \citet{Majewski03}.  Such a selection favors
more metal-rich populations.  Other samples such as SDSS SEGUE
spectra, favor metal-poor populations (see Appendix \ref{s:selfx}
below).  2) The definition of ``halo'' varies from author to author.
In many cases metallicity alone is used to define the halo.  3) The
volume probed can vary dramatically, from samples encompassing the
very local halo (e.g., within 1 kpc of the Sun), to sparse tracers
such as RR Lyrae that provide a view of the entire stellar halo.

\citet{Carollo07, Carollo10} used SDSS spectrophotometric calibration
stars to study the metallicity and orbital properties of the local
halo ($d<4$ kpc).  These stars were selected to have blue colors, and
therefore are strongly biased toward low metallicities, as noted by
the authors.  They identify two components to the stellar halo (which
they refer to as the ``dual halo''): a metal-rich ([Fe/H]$=-1.6$)
inner halo with highly eccentric orbits, and a metal-poor
([Fe/H]$=-2.2$) retrograde outer halo.  The transition between these
two components occurs around $R_{\rm gal}\approx 20$ kpc (the authors
were able to infer the properties of the halo beyond their $d<4$ kpc
selection by considering the maximum vertical extent of stars
throughout their orbits).  These authors fit multi-component Gaussians
to the MDFs in order to isolate various components.  We caution that
such a procedure can be difficult to interpret since MDFs for single
populations are expected on theoretical grounds to have a strong skew
toward low metallicities (e.g., closed box and other, more realistic
chemical evolution models; see Section \ref{s:fehprof}).  In agreement
with Carollo et al., we find that within $\sim30$ kpc the stellar halo
is dominated by a single population with highly radial orbits
\citep[referred to as {\it Gaia}-Enceladus, or Sausage in the recent
literature;][]{Helmi18, Belokurov18}.

However, in contrast with \citet{Carollo10}, at no distance or orbital
category do metal-poor stars (e.g., [Fe/H]$<-2$) dominate the
population.  We speculate that this difference is due to the fact that
Carollo et al. analyze stars within 4 kpc in order to infer the
properties of the halo at greater distances.  Any populations at large
distances that possess appreciable angular momentum will not be
well-represented in a local sample (the most striking example of this
is the Sagittarius stream, although the Sequoia remnant also possesses
a significant amount of angular momentum).

\citet{Liu18} use LAMOST spectra to study the MDF of A/F/G/K-type
stars with $|z|>5$ kpc.  By fitting Gaussians to the MDF, they identify
three distinct components with peaks of $-0.6$, $-1.2$ and $-2.0$; the
former they identify as the thick disk, and the latter two as the
inner and outer halo.  They show that the thick disk component is
confined to $|z|<10$ kpc, in broad agreement with our results.  They
also find that the retrograde stars are on average more metal-poor
than the prograde stars, also in agreement with our results.

Several studies have analyzed the global metallicity gradient of the
stellar halo.  Both \citet{Xue15} and \citet{Das16} used SEGUE K
giants to measure the metallicity gradient to $\approx100$ kpc.  They
define halo stars via a metallicity cut ($\feh<-1.2$ and $\feh<-1.4$,
respectively).  Both authors find evidence for a shallow (but
non-zero) metallicity gradient such that the metallicity decreases by
$\approx0.1-0.2$ dex from $10-100$ kpc.  In contrast, we find no
evidence for a metallicity gradient over the interval $\approx6-80$
kpc.  The differences are likely due to the metallicity cuts imposed
in the definition of the halo samples in \citet{Xue15} and
\citet{Das16}.  \citet{FernandezAlvar17} use APOGEE data to study the
halo metallicity profile.  They focus on stars with $|z|>5$ kpc that
satisfy a kinematic halo selection similar to what we employ.  They
find a flat gradient over the range
$10\lesssim R_{\rm gal}\lesssim 30$ kpc, in agreement with the results
presented here.

\citet{Xue15} estimate a mean metallicity of the halo of $\feh=-1.7$,
which is considerably more metal-poor than our value ($-1.2$).  There
are several reasons for this discrepancy.  First, Xue et al. {\it
  define} halo stars according to $\feh<-1.2$.  Second, the
metallicity scale of SEGUE appears to be slightly lower than H3 (see
Appendix \ref{s:comp}).  And third, the color selection used in the
SEGUE K giant sample excludes metal-rich stars (see Appendix
\ref{s:selfx}).

Recently, \citet{Mackereth19} use APOGEE data to isolate highly
eccentric stars on halo orbits and report an MDF that peaks at
[Fe/H]$\approx-1.3$.  They caution that the APOGEE selection function
makes it difficult to interpret the overall shape of the MDF.
Nonetheless, their MDF is in good agreement with our results.

In general, while there is broad agreement in the literature on the
main characteristics of the stellar halo, it is often difficult to
make quantitative comparisons owing to the fact that most previous
work has relied on metallicity-biased tracers of the halo, and/or have
focused on a small volume centered on the solar neighborhood.

\begin{figure}[!t]
\center
\includegraphics[width=0.47\textwidth]{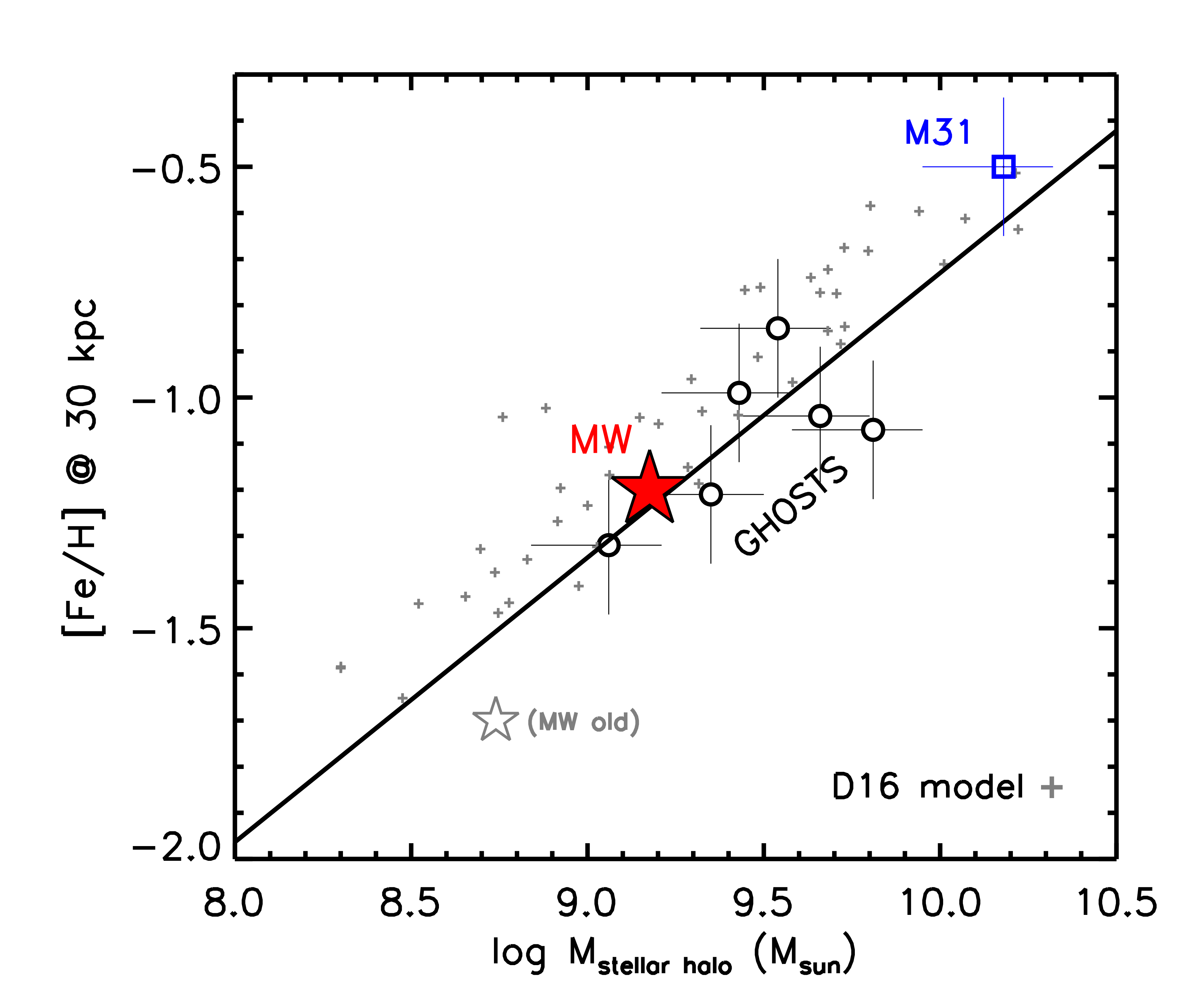}
\vspace{0.1cm}
\caption{Total stellar halo mass vs. stellar halo metallicity.  The
  metallicity is quoted at a common galactocentric distance of 30 kpc
  (although note that the MW gradient is flat so the choice of
  reference point will not change the location of the MW in this
  diagram).  The MW metallicity is from the present analysis and the
  adopted stellar halo mass is from \citet{Deason19}.  We also show
  the previous canonical values for the MW halo as a grey star.
  GHOSTS data are from \citet{Harmsen17} and the sources of the M31
  data are described in the text. Small grey symbols are predictions
  from the semi-empirical model of \citet{Deason16} and the solid line
  is a linear fit to the combined GHOSTS+M31+MW data.}
\label{fig:halofeh}
\end{figure}

\subsection{The Origin of the Stellar Halo}

A basic prediction of cold dark matter cosmology is the hierarchical
assembly of galaxies and their stellar halos
\citep[e.g.,][]{Johnston96, Helmi99, Bullock05}.  Evidence for the
tidal disruption of smaller dwarf galaxies is now ubiquitous both in
our Galaxy \citep[e.g.,][]{Majewski03} and beyond
\citep[e.g.,][]{Ibata01}.  Attempts to provide objective measures of
the degree of structure in the halo have found good agreement with
cosmological models \citep{Bell08}.

With a high-quality, unbiased sample of 4232 giants with well-measured
distances, proper motions, metallicities and abundances extending from
$6-100$ kpc, we are in a position to provide a holistic view of the
stellar halo.  This view is provisional for the reasons mentioned in
Section \ref{s:lim}, and will be updated as additional data are
collected.

The stellar halo beyond $|z|\gtrsim10$ kpc is overwhelmingly of
accreted origin.  We identified a population of stars with thick disk
chemistry that we associate with the in-situ halo.  Such stars
comprise $\approx25$\% of the halo at
$6\lesssim R_{\rm gal} \lesssim 10$ kpc, and only a few percent at
greater distances.  Our data do not probe halo stars at
$R_{\rm gal}<6$ kpc so it is possible that in-situ halo stars comprise
a greater fraction of the halo nearer to the Galactic center.  These
results are in broad agreement with predicted in-situ halo fractions
from the hydrodynamical simulations of \citet{Zolotov09}, who
predicted a large fraction on in-situ stars confined to the inner
regions of their simulated galaxies.  These results confirm and extend
previous analysis of {\it Gaia} DR1 data of the local stellar halo
(within 3 kpc of the Sun) by \citet{Bonaca17}.  These authors
identified a relatively-metal-rich population of halo stars with
thick-disk chemistry that they identified with the in-situ stellar
halo.

Previous work has identified at least four major chemical-orbital
structures in the halo: {\it Gaia-}Enceladus \cite{Helmi18,
  Belokurov18}, Sequoia \citep{Myeong19a}, Sagittarius
\citep{Majewski03}, and a metal-poor retrograde component
\citep{Carollo07, Helmi17}.  These four components are clearly visible
in our data.  The first three have remarkably similar average
metallicities (in the range $-1.0$ to $-1.3$), which helps to explain
the very flat metallicity gradient from $6-100$ kpc in spite of the
fact that different components dominate in different radial ranges.
The mass-metallicity relation at $z=0$ has a slope of 0.3 dex
\citep{Kirby13} and the dominant three components of the Milky Way
halo have estimated stellar masses that differ by a factor of
$10-100$.  One might therefore have expected a larger range in
metallicities.  However, the mass-metallicity relation is believed to
evolve with redshift, such that the zero point decreases with
increasing redshift \citep[e.g.,][]{Zahid13, Ma16}.  If the more
massive systems accreted earlier, then the evolving mass-metallicity
relation would result in a small range in metallicities amongst the
major remnants in the halo.

The nature of the metal-poor component ($\feh<-2$) is difficult to
discern based on the analysis presented here.  This population is
clearly distinct from the more metal-rich stars at $R_{\rm gal}>30$
kpc.  Such stars are also present in appreciable numbers at
$R_{\rm gal}<30$ kpc, and while they are not an obviously distinct
population based on metallicities alone, they do appear to occupy
distinct regions of orbital parameter space.  We conjecture that the
metal-poor component may in fact be tracing multiple distinct
populations.  This issue is discussed further in Carter et al. (in prep).

In summary, the data strongly favor a multi-component stellar halo
comprised primarily of accreted stars from at least four distinct
progenitor systems.  We expect additional structures to be identified
as new data allow sensitivity to lower-mass progenitor systems.

\subsection{The Galactic Stellar Halo In Context}

Several authors have explored the correlation between stellar halo
mass and metallicity both in observations and simulations.
\citet{Deason16} developed a semi-empirical model that predicted a
strong correlation between stellar halo mass and metallicity.  In
their model, this relation is set by the hierarchical assembly of dark
matter halos in conjunction with an empircally-constrained,
redshift-dependent stellar mass--halo mass relation and an empirical
mass--metallicity relation.  \citet{DSouza18b} and \citet{Monachesi19}
presented similar correlations based on the Illustris and Auriga
hydrodynamical simulations.  These authors compared their models to
observations of the stellar halos of the Milky Way (MW), M31, and six
galaxies from the GHOSTS Survey \citep{Harmsen17}.

The consensus from these comparisons is that the MW stellar halo has a
metallicity lower than expected for its mass.  In these comparisons a
stellar halo metallicity of $-1.6$ to $-1.7$ was adopted, along with a
stellar halo mass of $0.5\times10^9\Msun$.  Recently \citet{Deason19}
revised the stellar mass in the MW halo upward to
$1.5\times10^9\Msun$.  This significant upward revision results in a
MW stellar halo that is even more discrepant with the observed stellar
mass -- metallicity relation defined by other galaxies.  

One of the key results of our work is the higher average metallicity
of the Galactic stellar halo compared to previous work.  We therefore
revisit this issue in Figure \ref{fig:halofeh}, where we plot the
total stellar halo mass as a function of halo metallicity.  For the
MW, we adopt our average metallicity of $-1.2$, and the updated halo
mass from \citet{Deason19}.  The GHOSTS data are from
\citet{Harmsen17}, where the metallicities are quoted at 30 kpc.  The
M31 stellar halo mass is adopted from \citet{Harmsen17}, which is in
turn based on \citet{Ibata14}.  For the halo metallicity of M31 at 30
kpc, we follow \citet{DSouza18b} and adopt $\feh=-0.5$.  We also
include older estimates for the MW stellar halo mass
\citep{Bland-Hawthorn16} and metallicity \citep{Xue15}.

The revised stellar mass and metallicity of the halo places the Galaxy
on the locus defined by other galaxies.  NGC 4565 from the GHOSTS
Survey has a halo most closely resembling that of the Galaxy.
\citet{Harmsen17} quotes a total stellar mass for NGC 4565 of
$8\pm2\times10^{10}\Msun$, in broad agreement with modern estimates of
the total stellar mass of the Galaxy of $5-6\times10^{10}\Msun$
\citep{Licquia15, Bland-Hawthorn16}.  NGC 4565, an edge-on spiral
galaxy, may therefore be a useful Milky Way analog.


\section{Summary}
\label{s:sum}

In this paper we studied the stellar halo of the Galaxy using data
from the H3 Survey.  H3 selects targets based solely on {\it Gaia}
parallaxes and a magnitude cut, which produces the least biased view
of the stellar halo to-date.  We focused on a sample of 4232
kinematically-selected halo giants and presented the orbital and
chemical properties of halo stars to $\approx 100$ kpc.  Our key
findings are listed below.

\begin{itemize}

\item The stellar halo has a mean metallicity of
  $\langle\feh\rangle\approx-1.2$ with no discernable gradient from
  $6-100$ kpc.  Systematic uncertainties in the metallicity scale
  suggest that the mean metallicity could be as low as $-1.3$; lower
  metallicities are strongly disfavored.  The mean metallicity
  reported here is a significant upward revision in the mean halo
  metallicity, and is the result of the unbiased selection of
  spectroscopic targets in the H3 Survey.

\item This upward revision in the mean metallicity of the halo,
  combined with the recent upward revision in the total stellar mass
  of the halo by \citet{Deason19}, places the Galactic halo squarely
  in line with observations of other stellar halos in nearby galaxies.
  The Galactic halo metallicity is typical for its mass.  This higher
  mean metallicity also alleviates tension between the
  mass-metallicity relation and recent results favoring a single
  dominant progenitor contributing to the halo with a stellar mass of
  $\sim10^9\Msun$ \citep{Helmi18, Belokurov18}.

\item The stellar halo is rich in structure in chemical-orbital space,
  as predicted by hierarchical cosmological models.  We clearly
  identify a component of the halo with thick disk chemistry that is
  confined to $|z|\lesssim 10$ kpc, which we identify as the in-situ
  stellar halo.  Within $\approx30$ kpc the halo is dominated by stars
  on radial orbits and with a mean metallicity of $\feh=-1.2$.  We
  associate this with the {\it Gaia}-Enceladus merger remnant
  \citep{Helmi18}.  At greater distances the halo is comprised of at
  least two components: a prograde metal-rich component that is the
  Sagittarius stream, and a retrograde component slightly more
  metal-poor than the global mean.  This last component is likely
  associated with the Sequoia merger remnant \citep{Myeong19a}.  There
  is some evidence that the most metal-poor stars ($\feh<-2$) are a
  distinct component; detailed investigation of this possibility is
  the subject of ongoing work.

\item The picture emerging from these data is a stellar halo formed
  predominantly from the accretion and tidal disruption of multiple
  dwarf galaxies over cosmic time.  The inner halo contains a modest
  contribution from disk stars subsequently heated to halo-like orbits
  (the in-situ halo).  

\end{itemize}

Ongoing work is exploring the nature of these structures in greater
detail in chemical-orbital space along with comparisons to predictions
from models (Naidu et al. in prep; Carter et al. in prep; Johnson et
al. in prep).  The full H3 Survey dataset will more than double the
current sample size and survey footprint.  The final dataset will
therefore more than double the sensitivity to low-mass structures and
will deliver a more spatially complete view of the Galactic halo.


\acknowledgments 

We thank Eric Bell and Alis Deason for providing data in electronic
format.  We thank the Hectochelle operators Chun Ly, ShiAnne Kattner,
Perry Berlind, and Mike Calkins, and the CfA and U. Arizona TACs for
their continued support of the H3 Survey.  Observations reported here
were obtained at the MMT Observatory, a joint facility of the
Smithsonian Institution and the University of Arizona.


\begin{appendix}

\section{Comparison to Literature Metallicities}
\label{s:comp}

\begin{figure*}[!t]
\center
\includegraphics[width=0.95\textwidth]{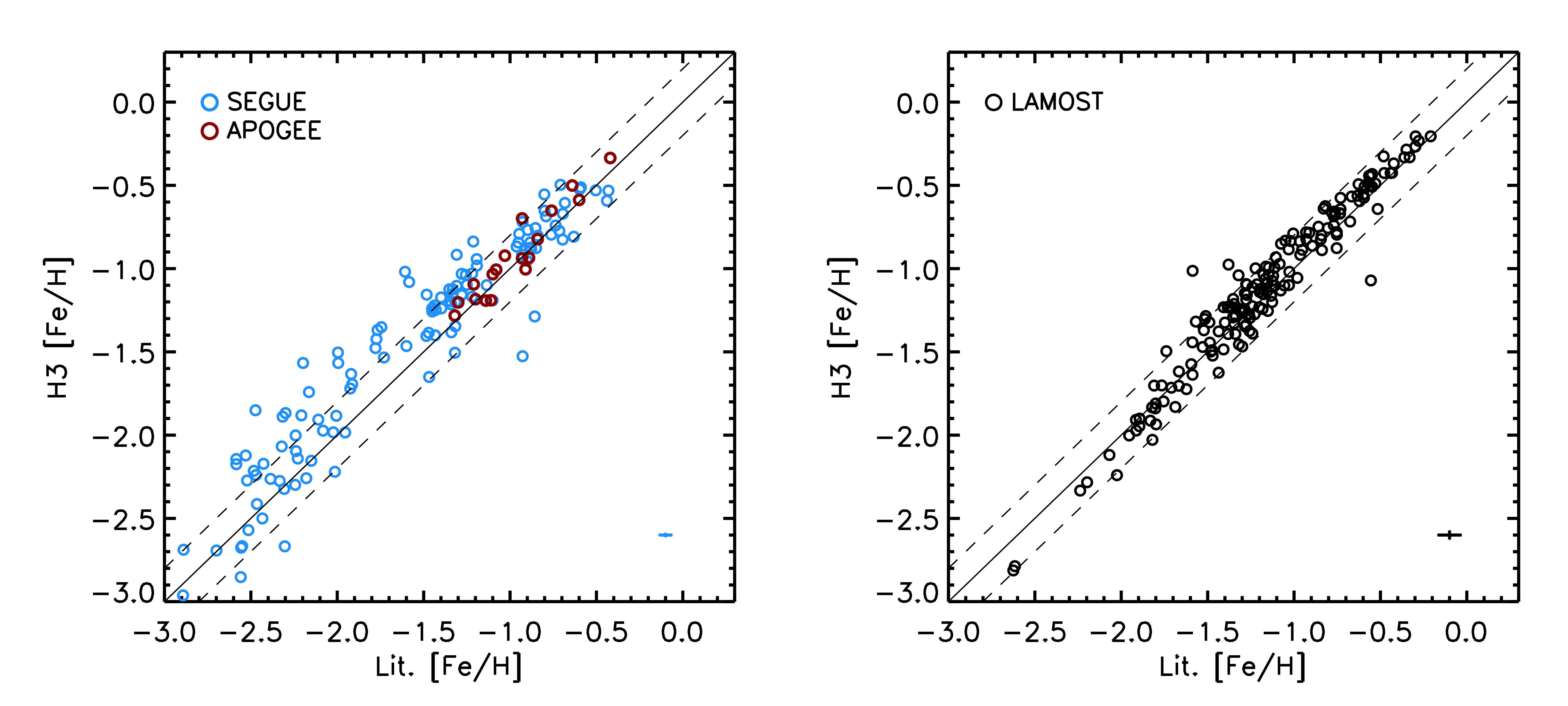}
\vspace{0.1cm}
\caption{Comparison between literature and H3 metallicities for giants
  $\logg<3.5$.  {\it Left Panel:} comparison between APOGEE, SEGUE,
  and H3 metallicities.  A $1-1$ line is shown for comparison, along
  with dashed lines offset by $\pm0.2$ dex.  Overall the agreement is
  good, although there is some evidence for a mild offset between
  SEGUE and H3 at $-2<\feh<-1$.  {\it Right Panel:} comparison between
  LAMOST and H3 metallicities.  In this case agreement is good across
  the entire metallicity range.  For SEGUE and LAMOST, typical error
  bars are shown in the lower right corner of each panel.}
\label{fig:lit}
\end{figure*}

In this section we compare the derived metallicities of the H3 data to
three independent large spectroscopic stellar surveys: SEGUE, APOGEE,
and LAMOST.  For SEGUE we use the SSPP parameters from DR14
\citep{LeeYS08, Smolinski11}, for APOGEE we use the parameters derived
in \citet{Ting19}, and for LAMOST we use parameters derived in
\citet{Xiang19}.  We have cross-matched the H3 catalog with public
catalogs of these three surveys.  We focus on giants with $\logg<3.5$,
and apply various quality flags where relevant.  For LAMOST we also
require SNR$_g>30$.  This results in 168, 121, and 18 stars with
$\logg<3.5$ in common between H3-LAMOST, H3-SEGUE, and H3-APOGEE.

In Figure \ref{fig:lit} we compare the metallicities for these stars
in common across the different surveys.  The left panel compares H3 to
SEGUE and APOGEE, while the right panel compares to LAMOST.  The
agreement between H3 and APOGEE is excellent, but the small overlap
between the samples limits the comparison to
$-1.5\lesssim\feh\lesssim-0.5$.  For LAMOST there are many more stars
in common and so the comparison extends over a much wider range in
metallicities.  Here the agreement is quite good over the entire
range.  At intermediate metallicities ($-1.5\lesssim\feh\lesssim-1.0$)
there is an approximately 0.1 dex offset between H3 and LAMOST such
that the former are more metal-rich.  At lower metallicities the sign
of the offset reverses such that H3 is approximately 0.1 dex more
metal-poor than LAMOST.  The LAMOST abundance scale in \citet{Xiang19}
is tied to APOGEE via a data-driven model, so the agreement between
LAMOST and H3 at some level guarantees good agreement also between H3
and APOGEE.

The comparison between H3 and SEGUE shows a more complicated picture
in part because of the sizable scatter between the two metallicity
estimates.  At $\feh>-1$ the agreement is overall quite good. There is
however some evidence for an offset in the range
$-2\lesssim\feh\lesssim-1$ between SEGUE and H3 such that H3
metallicities are $\approx0.2$ dex higher.  

This offset between SEGUE and H3 is puzzling because both surveys have
validated their stellar parameter pipelines against globular clusters
with low metallicities.  Focusing on clusters in the
$-2\lesssim\feh\lesssim-1$ range, \citet{Cargile19} demonstrated that
the H3 pipeline recovers metallicities for M13, M3, and M107 within
$0.1$ dex.  Specifically, for M13 they find $\feh=-1.47$ compared to
literature estimates ranging from $-1.50$ to $-1.53$.  For M3 they
find $\feh=-1.34$ compared to literature values of $-1.40$ to $-1.50$.
And for M107 they find $\feh=-0.92$ compared to a literature value of
$-1.01$.  For SEGUE data, \citet{LeeYS08} showed that their pipeline
recovers metallicities for M2 and M13 of $\feh=-1.52$ and $-1.59$
compared to literature values of $-1.62$ and $-1.54$.

It is beyond the scope of this paper to attempt to resolve the
tension.  We note however that the offset in [Fe/H] is correlated with
offsets in the derived [$\alpha$/Fe] values from each survey.  The
offset between SEGUE and H3 at low metallicity partially resolves the
discrepancy between the mean halo metallicity estimated in this work
compared to previous papers based on SEGUE data.  An additional factor
that explains some of the offset between SEGUE and H3 is discussed in
Appendix \ref{s:selfx}.

\section{Selection Effects in the SEGUE K-Giant Sample}
\label{s:selfx}

Prior to {\it Gaia} DR2, one of the most efficient means for selecting
candidate halo stars for spectroscopic follow-up was through
color-cuts designed to identify metal-poor stars.  An example of this
approach is the SDSS SEGUE spectroscopic sample of stars.
\citet{Yanny09} describe a variety of target selection types meant to
identify particular categories of stars.  In the context of stellar
halo science, one of the most popular has been the K giant target
types.  These samples were selected via a set of color cuts, including
the ``$l-$color'', defined as
$l=-0.436u+1.129g-0.119r-0.574i+0.1984$, where $ugri$ are
de-reddened SDSS magnitudes.

One must exercise caution when using samples defined according to a
series of color-based selections as these are likely to impart a bias
in the final sample.   In this Appendix we use the unbiased H3 data to
explore the impact of the l-color cut on the derived MDF.

\begin{figure}[!t]
\center
\includegraphics[width=0.47\textwidth]{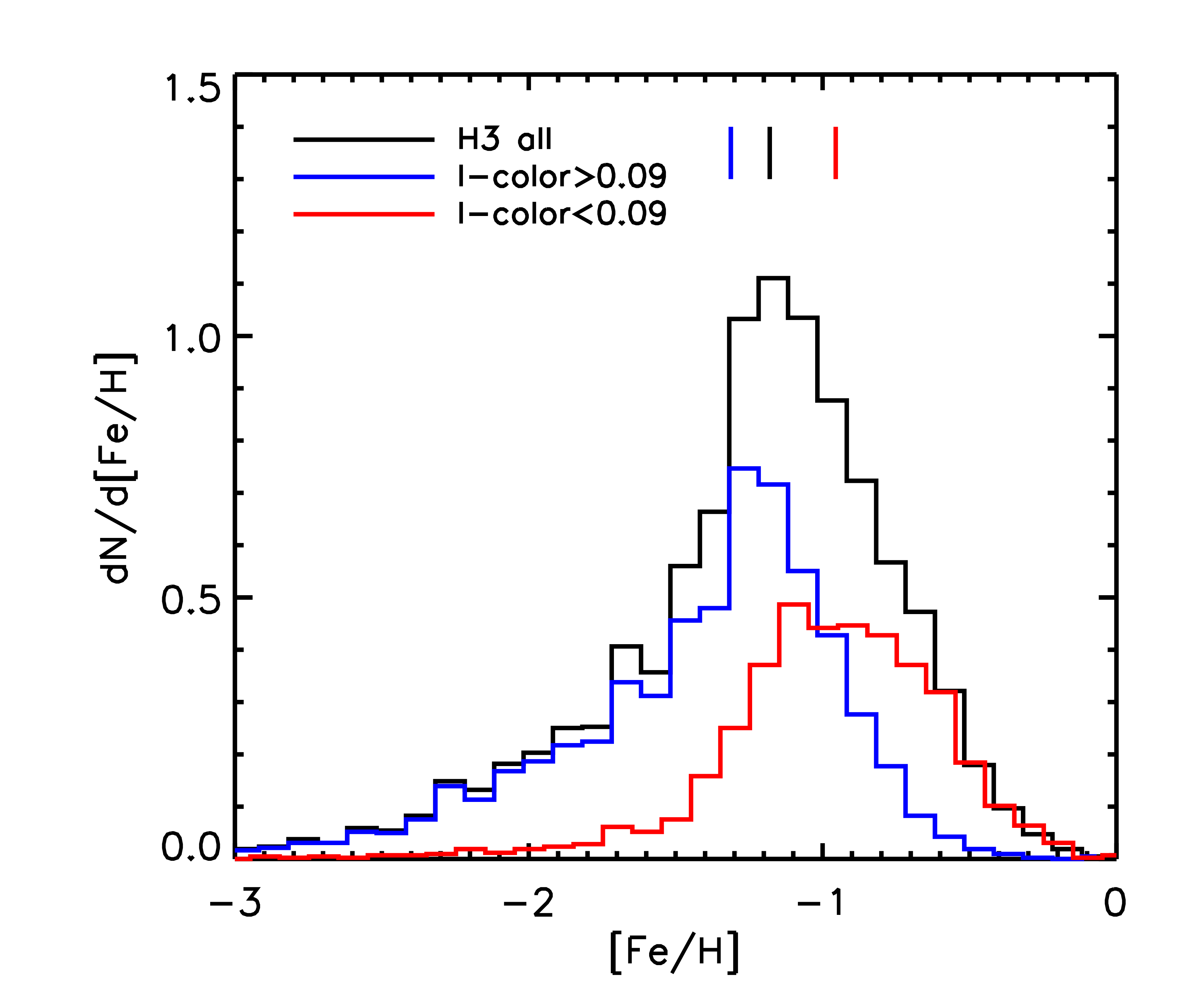}
\vspace{0.1cm}
\caption{Effect of the SEGUE $l-$color selection on the MDF.  The
  overall sample of kinematic halo giants from H3 (black line) is
  compared to subsamples where l-color$>0.09$ (blue line) and
  $l-$color$<0.09$ (red line).  The majority of K giants in the SEGUE
  sample were selected to have $l-$color$>0.09$, which clearly
  imprints a significant bias toward metal-poor populations. Vertical
  lines mark the median metallicities for each sample.}
\label{fig:lcol}
\end{figure}

Figure \ref{fig:lcol} shows the MDF for the H3 Survey, restricted to
the kinematically-selected halo giants.  The overall MDF is compared
to MDFs derived when adopting $l-$color$>0.09$ and $l-$color$<0.09$,
which is the main selection adopted by SEGUE-2 to identify K giants
\citep{Xue15}.  This figure demonstrates that this particular color
cut imposes a significant bias against the most metal-rich halo stars.

\end{appendix}


\end{document}